\documentclass[aps,pra,twocolumn,superscriptaddress,showpacs,showkeys,amsmath,fleqn]{revtex4}

\usepackage{graphicx}

\begin{document}

\title{Entanglement Swapping of Generalized Cat States and Secret Sharing}

\author{Vahid Karimipour}
\email[]{vahid@sina.sharif.edu}
\affiliation{Department of Physics, Sharif University of Technology\\
P.O. Box 11365-9161, Tehran, Iran}

\author{Saber Bagherinezhad}
\email[]{bagherin@ce.sharif.edu}
\affiliation{Department of Computer Science, Sharif University of Technology}

\author{Alireza Bahraminasab}
\email[]{bahramina@physics.sharif.edu}
\affiliation{Department of Physics, Sharif University of Technology\\
P.O. Box 11365-9161, Tehran, Iran}

\date{\today}

\begin{abstract}
We introduce generalized cat states for $d$-level systems and obtain
concise formulas for their entanglement swapping with generalized
Bell states. We then use this to provide both a generalization to
the  $d$-level case and a transparent proof of validity for an
already proposed protocol of secret sharing based on entanglement
swapping.
\end{abstract}


\maketitle

\section{Introduction}
There are numerous uses of spatially separated entangled pairs of
particles such as quantum key distribution and secret sharing
\cite{deutch, cerf, cabello1, cabello2, chineese}, teleportation
\cite{teleport1, teleport2}, superdense coding \cite{superdense}, and cheating
bit commitment \cite{cheating1, cheating2}. It has been argued that three or
more spatially separated particles in an entangled state (such as
a GHZ or cat state \cite{ghz1, ghz2, ghz3, ghz4}) may have similar or even broader
applications. It is first essential to distinguish between GHZ
states and cat states. By an $n$-party cat state, we mean a highly
entangled state of $n$ particles, while by a GHZ state we mean
one, which contradicts an interpretation in terms of any local
hidden variable theory. Constructing the latter types of states
for general multi-level multi-particle systems is quite a
difficult task, although some general criteria have been outlined
for their identification \cite{cabelloghz}. On the other hand we
will show that one can easily define $n$-party cat states with
nice properties (i.e: entanglement swapping) which allows them to
be used in a secret sharing protocol and possibly in many other
communication protocols, although they may not be used for testing
non-locality properties of quantum mechanics.  More recently,
applications such as reducing communication complexity
\cite{communication1, communication2} quantum telecomputation
\cite{telecomputation}, and networked cryptographic conferencing
\cite{conferencing11, conferencing12, conferencing13, conferencing2} have also been
suggested as possible new applications of these multi-particle entangled
states.

For practical applications such as mentioned above, there has been
much interest in manipulating entangled states of many particles
\cite{zukowski1, zukowski2, vedral, zeilinger}. In particular it
has been shown that by appropriate Bell measurements, entanglement
can be swapped between different particles\cite{zukowski1}, a
scheme which has been generalized to multi particle case in
\cite{vedral}. In fact to the question of ``Which particles get
entangled when we make cat state measurement on a group of
particles?'', there is a general pencil and paper rule which
provides the answer \cite{vedral}. One just has to connect the
particles being measured to frame a polygon and those not being
measured to frame a complementary polygon. These two polygons
represent the two multi-particle cat states obtained after the
manipulation. However for most applications it is highly necessary
to know exactly the type (e.g. the labels) of the cat state that
the particles are forming and a knowledge of only the particles
sharing the entanglement is not enough. In fact in almost any of
the communication protocols mentioned above, the information to be
transferred is encoded in the type of the labels of the cat states
involved. For this reason one needs also a simple pencil and paper
rule for determining the types and the labels of the cat states
involved in a swapping process.

It may not be so illuminating to derive a general formula for such
a purpose, although it is rather straightforward to do so. However
if we restrict to the most common type of swapping, that is, the
swapping of a cat state and a Bell state, then transparent,
graphical and very useful rules can be derived as we will show
below. Furthermore we will derive the rules for general $d$-level
systems. We will then apply these rules to the quantum key
distribution and secret sharing protocols of  \cite{cabello1,
cabello2} and show that the rules of encoding and decoding of this
protocol, expressed otherwise only in tables, even when few
parties are involved \cite{cabello2} can be neatly expressed by
closed formulas in the general case.

The structure of this paper
is as follows. In section 2 we review the basic properties of
$d$-level Bell states \cite{bell1, bell2, bell3} and introduce $d$-level cat states.
In section 3 we derive simple graphical rules for entanglement
swapping of $d$-level Bell and cat states. We then apply in section
4, these rules to the secret sharing protocol of Cabello
\cite{cabello2} to see how simple the encoding and decoding rules
of this protocol are. We conclude the paper with a discussion.

\section{Generalized cat states for $d$-level systems}
In studying $d$-level states and their entanglement properties we
are following an interesting trend to generalize the well known
quantum algorithms and protocols of quantum computation and
communication to non-binary systems, like quantum gates for qudits
\cite{bartlett}, quantum error correcting codes \cite{knill,
chau}, and generalization of the BB84 protocol \cite{bb84} for
quantum key distribution\cite{cerf}.
(For a review on quantum key distribution see \cite{rev}.)

In fact considerations of quantum hardware may bring about some
advantage to non-binary systems, since bigger Hilbert spaces can
be made by coupling fewer d-dimensional systems than 2-dimensional
ones, and it is well known that complete coupling of quantum bits
gets much more difficult with the number of qubits increasing.
Some researchers have even considered quantum computation and
communication with continuous variables \cite{cont1, cont2}.
Besides these, it is very instructive to study quantum computation
and communication for $d$-level
systems (qudits) to understand them in a general dimension-free setting.

\begin{figure*}
\includegraphics{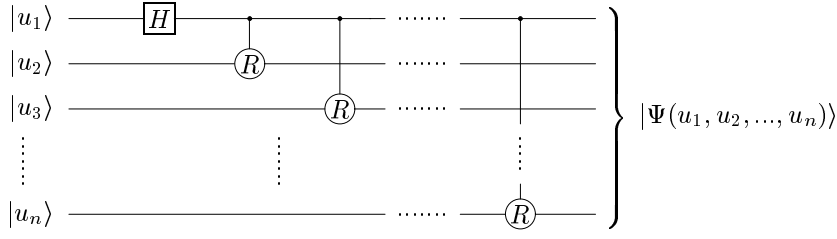}
\caption{\label{fig:circuit} Circuit for constructing $d$-level cat states.}
\end{figure*}

We start by  reviewing  a generalization of the familiar Bell
states for qudits introduced in \cite{bell1, bell2, bell3}. These are a set of
$d^2$ maximally entangled states which form an orthonormal basis
for the space of two qudits. Their explicit forms are:
\begin{equation}\label{bell}
 |{\bf \Psi}(u_1, u_2)\rangle := \frac{1}{\sqrt{d}}\sum_{j=0}^{d-1}
  \zeta^{j u_1}|j, j+u_2\rangle
\end{equation}
where $ \zeta = e^\frac{2\pi i}{d}$ and $u_1 $ and $ u_2 $ run from $
0 $ to $ d-1$. Each Bell state is thus characterized by a pair of
two $ Z_d $ labels. For $d=2$ these states reduce to the familar
Bell states, usually denoted by $ |{\bf \Psi}^{\pm} \rangle $ and $
|\Phi^{\pm} \rangle $. One can also expand any computational basis
vector in terms of Bell states:
\begin{equation}\label{inversebell}
  |j, k\rangle = \frac{1}{\sqrt{d}}\sum_{u=0}^{d-1} \zeta^{ - j u} |{\bf \Psi} (u,k-j)\rangle
.
\end{equation}
It is also useful to consider a generalization of the familiar Hadamard gate to the
$d$-level case.
It is defined \cite{mermin, kbb} as follows:
\begin{equation}\label{hadamardket}
  H =\frac{1}{\sqrt{d}} \sum_{i,j=0}^{d-1} \zeta^{ij}|i\rangle\langle j|.
\end{equation}
This operator is really not new and it is known as the quantum
fourier transform when $d = 2 ^n$. In that case it acts on $ n $
qubits. Here we are assuming it to be a basic gate on one single
qudit, in the same way that the ordinary Hadamard gate is a basic
gate on one qubit. It is also useful to generalize the NOT and the
CNOT gates. We note that in the context of qubits, the NOT gate,
is basically a mod-2 adder. For qudits this operator gives way to
a mod-$d$ adder, or a Right-Shift gate.

\begin{equation}\label{rightandleftshift}
  R|j\rangle = |j+1\rangle\text{mod} d,
\end{equation}
where here and hereafter all our additions are defined mod $ d $.
Note that $ R^d = I$, compared to $ NOT^2 = I$. For any unitary
operator $ U $, the controlled operator $ U_c $ is naturally
generalized as follows:
\begin{equation}\label{controlledgate}
  U_{c}(|i\rangle\otimes|j\rangle) = |i\rangle\otimes U^i|j\rangle
\end{equation}
Here the first and the second qudits are respectively the
controller and the target qudits. In  particular the controlled
shift gates which play the role of CNOT gate, act as follows:
\begin{equation}\label{controlledshift}
  R_c |i,j\rangle = |i,j+i\rangle
\end{equation}
Equipped with the $d$-level Hadamard and CNOT ($R_c$) gates, one can
construct $d$-level cat states simply as in the 2-level case by the
circuit shown in fig. (\ref{fig:circuit}), where $|u_1, u_2, \cdots, u_n\rangle $ is
a computational basis vector, in which $ u_i \in \{ 0, 1, 2,
\cdots, d-1\}$.

The resulting cat state will be:
\begin{eqnarray}
|{\bf \Psi}(u_1, u_2, \cdots, u_n)\rangle:=\nonumber\\
\frac{1}{\sqrt{d}}\sum_{j=0}^{d-1} \zeta^{j u_1}|j, j+u_2,
j+u_3, \cdots, j+u_n\rangle.\label{cat}
\end{eqnarray}
These states are orthonormal, $ \langle {\bf \Psi}(v_1, \cdots,
 v_n)|{\bf \Psi}(u_1, \cdots, u_n)\rangle = \delta_{u_1, v_1} \cdots
\delta_{u_n, v_n}$ and complete: any computational basis vector
can be expanded in terms of these generalized cat states:
\begin{eqnarray}
|u_1, u_2, u_3, \cdots, u_n \rangle= \nonumber\\
\frac{1}{\sqrt{d}}\sum_{j=0}^{d-1} \zeta^{-j u_1} |{\bf
\Psi}(j, u_2-u_1, u_3- u_1, \cdots, u_n- u_1)\rangle,\label{inversecat}
\end{eqnarray}
Quite analogously to the 2-level case, one can generate a cat
state of $n$ particles from a cat state of $n-1$ particles in two
ways, either using Zeilinger et. al. method \cite{zeilinger}, that
is: acting by an $R_c$ gate on one particle of the $(n-1)$-cat state
and one qudit of a Bell state, subsequently measuring the target
qudit, or by using the method of \cite {vedral} by performing a
Bell state measurement on two particles, one from an $(n-1)$-cat
state and the other from a 3-cat state, projecting the rest
onto an $n$-cat state.

\section{Some Simple Rules for Entanglement Swapping}

\begin{figure}
\includegraphics{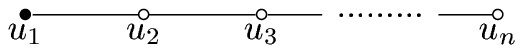}
\caption{\label{fig:cat} Visualization of $|{\bf \Psi}(u_1, u_2, \cdots, u_n)\rangle$ cat
state.}
\end{figure}

Entanglement swapping is nothing but tensor multiplying two cat
states, expanding them in the computational basis of the product
space, swapping a subset of particles and then re-expanding the
resulting state in terms of the new cat states. The idea and the
essential calculation is best illustrated by the simplest example,
that is, swapping two Bell states. Suppose particles 1 and 2 are
in a Bell state $|{\bf \Psi}(u_1, u_2)\rangle_{_{1,2}} $ and particles
3 and 4 are in a Bell state $|{\bf \Psi}(v_1, v_2)\rangle_{_{3,4}} $.
This state of the four particles is equal to:
\begin{eqnarray}
 \frac{1}{d} \sum_{j,j'} \zeta^{j u_1+j'v_1}
|j, j+u_2\rangle_{_{1,2}} |j',j'+v_2\rangle_{_{3,4}}&=& \nonumber\\
\frac{1}{d} \sum_{j,j'} \zeta^{j u_1+j'v_1}
|j, j'+v_2\rangle_{_{1,4}} |j',j+u_2\rangle_{_{3,2}}&=& \nonumber\\
\frac{1}{d^2} \sum_{j, j',w, w'} \zeta^{j u_1+j'v_1} \zeta^{-j
w -j'w'}|{\bf \Psi}(w, j'+v_2-j)\rangle_{_{1,4}}&&\nonumber\\
|{\bf \Psi}(w',j+u_2-j')\rangle_{_{3,2}}&&\label{detailedswap}
\end{eqnarray}
Changing the variables ($j'-j \rightarrow \ell $), and using the
identity $  \frac{1}{d}\sum_{j=0}^{d-1}\zeta^{j n} = \delta
(n,0)$, and rearranging terms we finally arrive at:
\begin{eqnarray}
|{\bf \Psi}(u_1, u_2)\rangle_{_{1,2}} |{\bf \Psi}(v_1,
v_2)\rangle_{_{3,4}}&=&\nonumber\\
\frac{1}{d} \sum_{k, \ell} \zeta^{-k \ell}|{\bf \Psi}(u_1 + k, v_2+\ell)\rangle_{_{1,4}}
|{\bf \Psi}(v_1- k, u_2-\ell)\rangle_{_{3,2}}&&\label{final}
\end{eqnarray}

\begin{figure}
\includegraphics{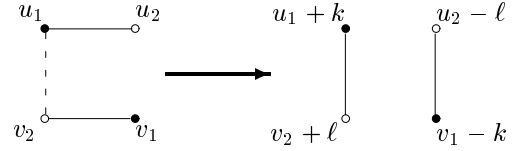}
\caption{\label{fig:Bell} Entanglement swapping of $d$-level Bell states.}
\end{figure}

It is customary to represent a cat state by a polygon. However a
cat state is not symmetric and a polygon can not represent it
properly. In fact as it is clear from \ref{cat} that a cat state is
symmetric under the interchange of both the labels and the
particles from $2$ to $n$, i.e:
\begin{eqnarray}
& & |{\bf \Psi}(u_1, \cdots, u_k, \cdots, u_l, \cdots,
u_n)\rangle{_{1,\cdots, k, \cdots, l, \cdots, n}}\nonumber\\ 
&=&|{\bf \Psi}(u_1,\cdots, u_l, \cdots, u_k, \cdots, u_n)\rangle_{_{1, \cdots, l, \cdots,
k, \cdots, n}}\label{symmetry}
\end{eqnarray}
however it has no such symmetry under the interchange of the first
particle with another one. We therefore depict a cat state by a
line with $n$ nodes on it, distinguishing the first node from the
others by by assigning a black circle to it compared with empty
circles assigned to others (fig. (\ref{fig:cat})). With this convention, the
result of swapping calculated in equation \ref{detailedswap} can
be depicted as in fig. (\ref{fig:Bell}), where we have ignored the coefficients
of the expansion and the arrow is meant to imply that the right
hand side is a possible outcome of the Bell measurement performed
on the left hand side particles designated by dashed line. The
simple rule is that the sum of labels on the black nodes and white
nodes are conserved separately in such a swapping. We will see
that this type of rule will also hold true with slight
modifications in swapping of Bell states and cat states.

\begin{figure*}
\includegraphics{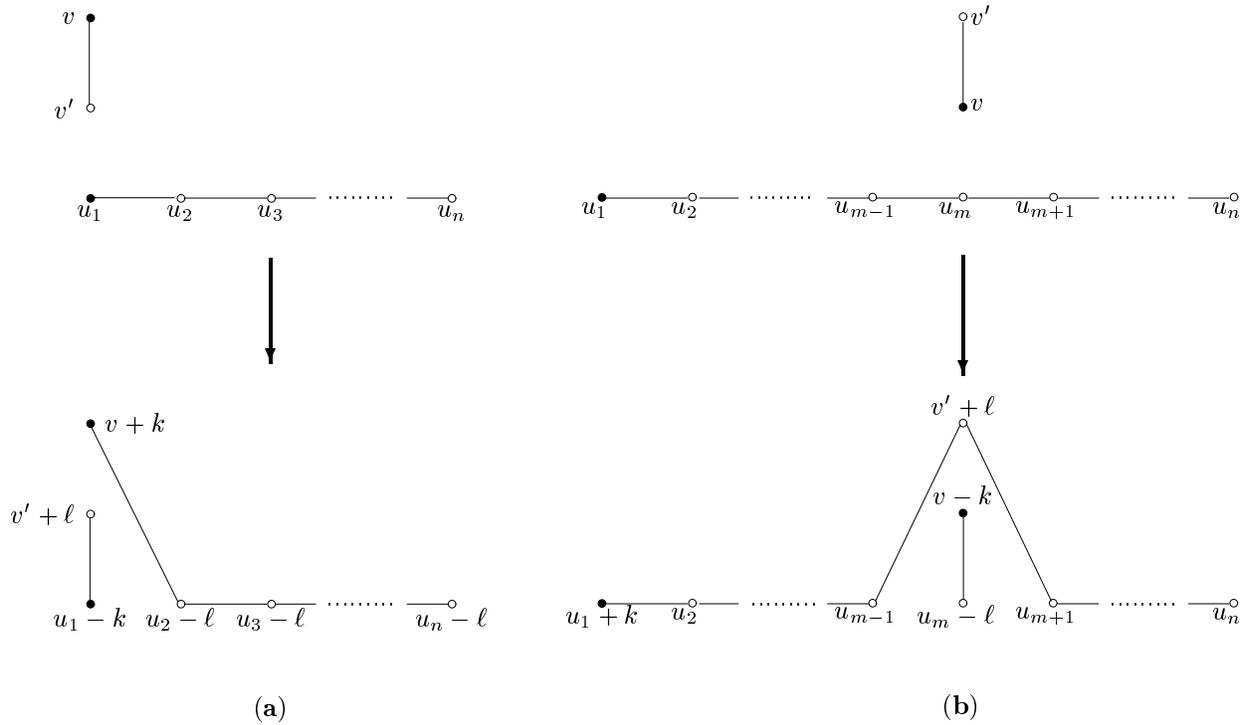}
\caption{\label{fig:catBell} Entanglement swapping between a cat state and a Bell state}
\end{figure*}

We now
derive formulas for swapping Bell states and cat states. We
distinguish two cases, one in which a Bell state measurement
involves the first particle (the black node) of the cat state and
one in which it does not. For the first case we find after some
straightforward calculations:
\begin{eqnarray}
|{\bf \Psi}(u_1, u_2, \cdots, u_n)\rangle _{_{1, 2, \cdots, n}}
\otimes|{\bf \Psi}(v, v') \rangle_{_{s,s'}} =&&\nonumber\\
\frac{1}{d}\sum_{k,\ell} \zeta^{-lk} |{\bf \Psi}(v+k, u_2-\ell, u_3-\ell,
\cdots, u_n-\ell)\rangle_{_{s,2,3,4,\cdots, n}}&\otimes&\nonumber\\
|{\bf\Psi}(u_1-k, v'+\ell)\rangle_{_{1,s'}}\label{swap1}
\end{eqnarray}
This formula is depicted graphically in fig. (\ref{fig:catBell}-{\bf a}). Again we see a
simple rule in terms of the conservation of the labels on the
black and white nodes. For the second case where the Bell state
measurement does not involve the black node of the cat state we
find:
\begin{eqnarray}
|{\bf \Psi}(u_1, u_2, \cdots, u_n)\rangle _{_{1,2,3,\cdots, n}}
\otimes|{\bf \Psi}(v, v') \rangle_{_{s,s'}} =&&\nonumber\\
\frac{1}{d}\sum_{k,\ell} \zeta^{-\ell k} |{\bf \Psi}(u_1+k, u_2, u_3,
\cdots, v' + \ell,\cdots, u_n )\rangle_{_{1,2,\cdots, s',\cdots,
n}}&\otimes& \nonumber\\ 
|{\bf \Psi}(v-k, u_m-\ell)\rangle_{_{s,m}}&&\label{swap2}
\end{eqnarray}
This is depicted in fig. (\ref{fig:catBell}-{\bf b}).

\section{Secret key sharing by entanglement swapping}
\begin{figure*}
\includegraphics{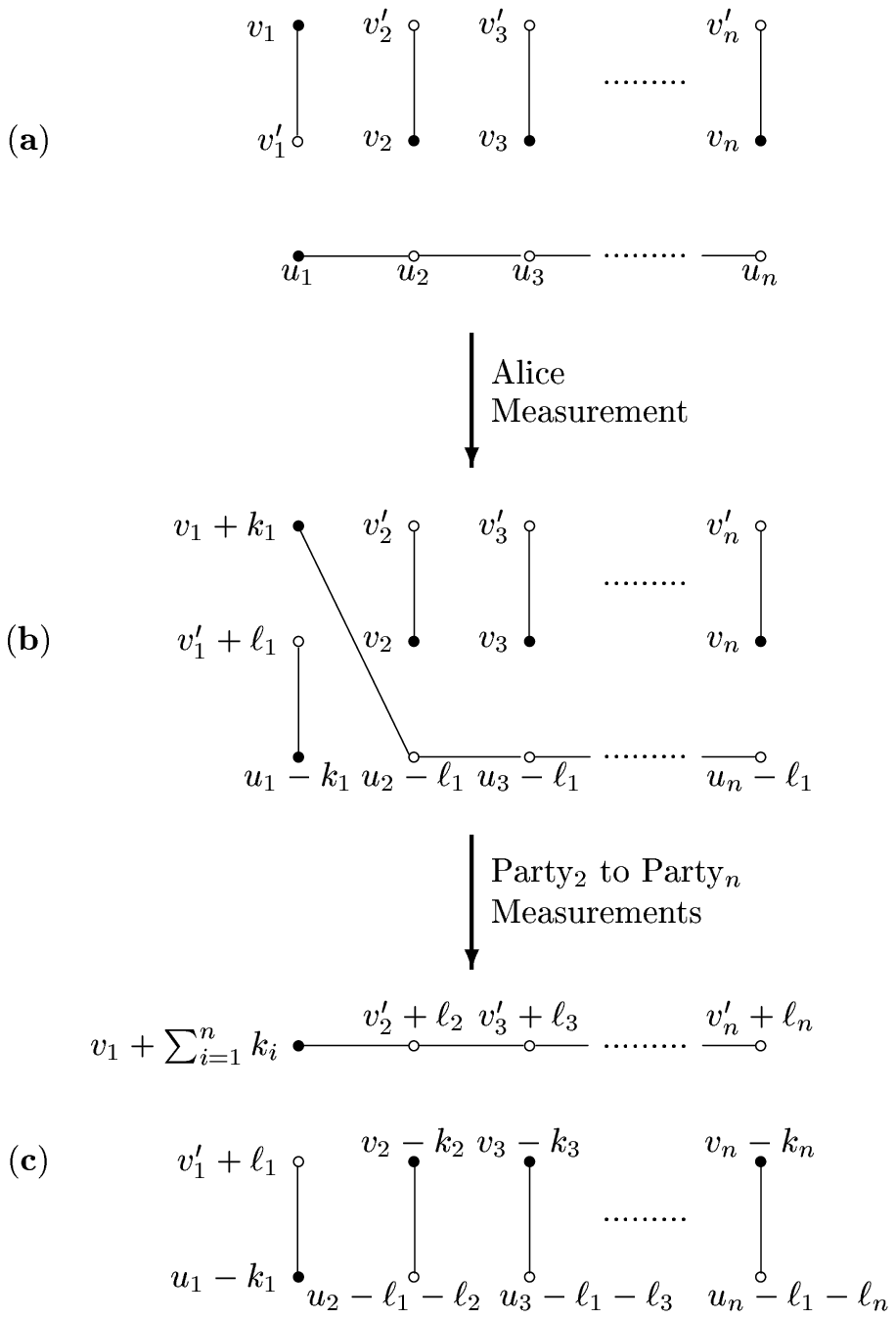}
\caption{\label{fig:Cabello} A protocol for $d$-level secret sharing.}
\end{figure*}

Among the many applications of entanglement swapping mentioned in
the introduction, in this section we consider the secret key
sharing protocol proposed by Cabello in \cite{cabello1, cabello2}.
In this protocol $n$ members of a group want to agree upon a
secret key (For $n=2$, we have the simple QKD scheme). The key is
to be such that no proper subset of the group can determine it and
its determination requires the cooperation of all members of the
group. In the protocol proposed by Cabello \cite{cabello2} the $n$
members of the group share an $n$-cat state, and each of them has
also a Bell state. Each of the members swaps her or his Bell state
with the cat state, and then all of them send the resulting cat
state to one of the members say Alice, who measures the cat state
and announces the result of her measurement in public. It is then
argued that by using this knowledge and the result of their own
Bell measurements, the members of the group can all determine the
result of the Bell measurement of Alice, which is to act as a
random two bit key. In \cite{cabello2}, it is shown by way of a
couple of examples for 3 and 4 parties and compiling the results
of measurements in tables, that this is indeed possible. Here we
generalize the results of \cite{cabello2} in two respects. First
we consider general $d$-level systems instead of two level ones,
second by using our simple rules for entanglement swapping we
derive general and concise formulas for determining the final
secret key in terms of measurements of individuals members. As
mentioned above, we can carry out all of the analysis graphically,
where by our graphics we not only imply the particles which get
entangled under swapping but also indicate precisely the entangled
states they form in this process. The first stage of the process
is depicted in fig. (\ref{fig:Cabello}-{\bf a}), where each member
(i), has a Bell state $(v_i, v'_i)$ and all the members share also
a cat state $ (u_1, u_2, \cdots, u_n)$.  When the first member
whom we call Alice, performs her Bell measurement the entanglement
swaps to the form shown in fig. (\ref{fig:Cabello}-{\bf b}), where
we have used the first rule of fig. (\ref{fig:catBell}-{\bf a}).
Subsequently members numbered 2, 3, ... and $n$ perform their Bell
measurement and the states swap to that of fig.
(\ref{fig:Cabello}-{\bf c}). The random two dit key is the pair of
labels of Alice's Bell state, that is $ ( u_1-k_1, v'_1 + \ell_1)$
. At this stage  the cat state is sent to Alice, she measures the
state and announces the labels $( v_1+ k_1 + k_2 + \cdots + k_n,
v'_2 + \ell_2, v'_3 + \ell_3, \cdots, v'_n + \ell_n)$ of this
state in public. It is now clear that each member of the group say
the $i$-th one, $(i = 2, 3, \cdots, n)$, knowing his own Bell
state $ (v_i, v'_i)$ at the beginning of the protocol, his final
Bell state $(v_i - k_i, u_i - \ell_1 - \ell_i)$  and the publicly
announced cat state,  can independently determine $ \ell_1 $ and
hence the second label of the secret key, $v'_1 + \ell_1$. (Note that
the shared cat state labels and all the Bell labels including
those of Alice $(v_1, v'_1)$ are assumed to be known to all the
members at the beginning of the protocol). However to determine
the first label of the key, that is $ u_1 - k_1 $, the members
need a knowledge of $ k_1 $, which no subset of the group can
determine independently. It can only be found by sharing their
values of $ k_i, i = 2, 3, \cdots, n$ with each other. Once this
is done all members can determine the value of $k_1$ from the
publicly announced label of Alice $v_1 + k_1 + k_2 + \cdots +
k_n$.

The way we have presented this protocol, which starts with general
Cat and Bell states, rather than with special ones with say all
the labels being zero (i.e. $\Phi(0,0)\rangle$), has the advantage
that it shows how the encoding and decoding scheme works for
consecutive qudits, when the same Bell and Cat states are re-used.
To compare our results with those of \cite{cabello2}, it is enough
to set all the original labels $ u_i,v_i,v'_i = 0 $. It is then
easy to see from fig. (\ref{fig:Cabello}) that our results
completely match the tables presented in that article.
\section{Discussion}
We have provided closed formulas for entanglement swapping of
$d$-level cat states and Bell states. We have then used our formulas
for providing transparent proof for the validity of a secret
sharing protocol between $n$ parties based on entanglement swapping.
We expect that our graphical method of representing cat states and
our formulas for entanglement swapping (ES) may find applications
in every ES-based protocol in quantum communication.

\bibliography{swapp}

\end{document}